\documentstyle[11pt,epsf,fleqn]{article}
\begin{document}
\title{Electroweak phase transition in the MSSM with four generations}
\author{S. W. Ham$^{(1)}$, S. K. Oh$^{(1,2)}$, D. Son$^{(1)}$,
\\
\\
{\it $^{(1)}$ Center for High Energy Physics, Kyungpook National University}\\
{\it Daegu 702-701, Korea} \\
{\it $^{(2)}$ Department of Physics, Konkuk University, Seoul 143-701, Korea}
\\
\\
}
\date{}
\maketitle
\begin{abstract}
By assuming the existence of the sequential fourth generation to the minimal supersymmetric
standard model (MSSM), we study the possibility of a strongly first-order electroweak phase transition.
We find that there is a parameter region of the MSSM where the electroweak phase transition is
strongly first order.
In that parameter region, the mass of the lighter scalar Higgs boson is calculated to be
above the experimental lower bound, and the scalar quarks of the third and
the fourth generations are heavier than the corresponding quarks.
\end{abstract}
\vfil
\eject

\section{Introduction}

As a mechanism to explain the baryon asymmetry of the universe, the electroweak baryogenesis
is given wide attentions, since it can be tested in the future high energy experiments [1].
Sakharov have several decades ago established the three essential conditions for generating
dynamically the baryon asymmetry of the universe from a baryon-symmetric universe [2].
As is well known, the three conditions are the presence of baryon number violation,
the violation of both C and CP, and a departure from thermal equilibrium.
The possibility of the electroweak phase transition has already been exhaustively studied,
which can provide the baryon number violation and the violation of both C and CP.
The remaining Sakharov condition, the departure from thermal equilibrium, may be
fulfilled at a weak scale temperatures if the nature of the electroweak phase transition is first order.
The difficulty of the standard model (SM) is that the strength of the first order electroweak
phase transition, which must be strong enough for preserving the generated baryon asymmetry
at the electroweak scale, appears too weak for the experimentally allowed mass of
the SM scalar Higgs boson [3].

Thus, it seems that electroweak baryogenesis requires a new physics beyond the SM at weak scale [4].
The minimal supersymmetric standard model (MSSM) has been studied intensively
within the context of electroweak baryogenesis.
It is observed that, if one of the scalar top quark has a mass smaller than top quark mass,
the MSSM may possess a parameter region where the electroweak phase transition is strong enough [5].
In this scenario of a light scalar top quark, the requirement that the electroweak phase transition
should be strongly first order is equivalent to the imposition of an upper bound of about 120 GeV
on the lightest Higgs scalar boson mass of the MSSM.

In any supersymmetric standard model, there are superpartners to ordinary quarks and leptons.
Experimentally, no scalar quark or scalar lepton that is lighter than 5 GeV has been discovered.
Thus, the ordinary quarks and leptons, up to bottom quark, are lighter than their corresponding
superpartners.
Possibly, top quark might be an exception, if a scalar top quark is lighter than top quark.

The idea that a scalar top quark might be lighter than top quark is not in accord
with the behavior of lighter fermions, but is allowed by present experiment
and accommodated in the MSSM within the context of the electroweak phase transition.
If the scalar top quarks are not degenerate in mass, the $2\times2$ mass matrix for
the scalar top quarks yields a lighter scalar top quark and a heavier one in the MSSM.
In order for the electroweak phase transition to be strongly first order,
the mass of the scalar top quark should be either smaller than 170 GeV or larger than 1 TeV.
Thus, the lighter scalar top quark should have a mass smaller than 170 GeV.
The heavier scalar top quark may have a mass comparable to the SUSY breaking scale, say,
between 1 and 2 TeV.
A light scalar top quark suggested in the MSSM scenario, which depends strongly on chargino
and neutralino masses, might soon face experimental examination at Tevatron.

The possibility of a scalar top quark heavier than top quark has been considered
in scenarios other than the MSSM. For example, in the next-to-minimal supersymmetric
standard model (NMSSM), the strongly first order electroweak phase transition takes place
where the top quark mass is smaller than the mass of the scalar top quark.
Within the framework of the MSSM, we need to enlarge the model somehow to accommodate
a scalar top quark heavier than top quark.
We examine the possibility for the strongly first-order electroweak phase transition
in the MSSM by introducing an extra generation of fermions.

In this paper, we study the effect of the fourth generation of quarks on the strength
of the first order electroweak phase transition in the MSSM.
We find that introducing an extra generation of fermions might also lead to a scalar top quark
heavier than top quark. Although the number of the SM neutrino species has already been
fixed experimentally as three, one may find a lot of articles in the literature
that mention about the fourth generation and study on the assumption of its existence [6,7].
In principle, the SM, as well as the MSSM, can accomodate any number of generations.
In various contexts, the MSSM with four generations has been studied [8,9,10].

Our study shows that the fourth generation is found to enhance the strength of the first-order
electroweak phase transition, while the scalar Higgs boson mass is calculated to be larger
than the experimental lower bound, and the scalar quarks of the third generations are comparable
to the supersymmetry breaking scale ($M_{\rm SUSY} = 1$ TeV),
in a reasonably wide region of parameter space in the MSSM.
In our scenario, a light scalar quark is not necessarily required to ensure
the first order electroweak phase transition be strong;
the scalar quarks of the third and the fourth generations may be heavier than the corresponding quarks;
and the scalar Higgs boson mass lies above the experimental lower bound.

\section{Higgs potential in decoupling limit without mixing}

Let us study a particular, yet reasonable as well as plausible,  form of the Higgs potential
in the MSSM with four generations of quarks for the electroweak phase transition.
We consider only the third and the fourth generations, and assume that there is no mixing between them.
The fourth generation appears simply in a repetitive manner.
As is well known, there are two Higgs doublets in the Higgs sector of the MSSM, namely,
$H_1^T = (H_1^0, H_1^-)$ and $H_2^T = (H_2^+, H_2^0)$.
After electroweak symmetry breaking emerge five physical Higgs bosons:
two neutral scalar Higgs bosons ($h \ , H$), one neutral pseudoscalar Higgs bosons ($A$),
and a pair of charged Higgs bosons ($H^{\pm}$). We assume that the mass of $h$
is lighter than that of $H$.
At the tree level, the Higgs sector of the MSSM depends on only two free parameters.
We take them to be the ratio $\tan \beta = v_2/v_1$ of the two real vacuum expectation values (VEVs)
$v_1$ of $H^0_1$ and $v_2$ of $H^0_2$ and $m_A$, the mass of $A$.
In this paper, we assume that CP is conserved in the Higgs sector by choosing all parameters
in the effective Higgs potential to be real.

In the decoupling limit, where $m_A \gg m_Z$, with fixed $\tan \beta$, only one linear combination of
the two neutral scalar Higgs bosons,
    \begin{equation}
    \phi = \sqrt{2} \cos \beta {\rm Re} (H_1^0) + \sqrt{2} \sin \beta {\rm Re} (H_2^0) \ ,
    \end{equation}
remains light at the electroweak scale [11].
In this limit, the tree-level Higgs potential at zero temperature can be expressed
in terms of $\phi$ as
    \begin{equation}
    V_0(\phi, 0) =  - m_0^2 \ \phi^2 + {\lambda \over 4} \phi^4 \ .
    \end{equation}
Since all quartic terms have gauge coupling coefficients in the MSSM,
the quartic Higgs self-coupling $\lambda$ is
given as $\lambda = (g_1^2 + g_2^2)/4$.
Note that there is an upper bound on the mass of $h$ as $m_h \le m_Z |\cos 2 \beta|$
at the tree level in the MSSM.
In this limit, the couplings of $h$ to gauge bosons and fermions are identical
to the couplings of the SM Higgs boson,
which implies that one cannot distinguish phenomenologically the SM scalar Higgs boson from $h$ [12].
Thus, one might expect that $m_h$ has the same experimental lower bound
as the SM scalar Higgs boson in the decoupling limit [13].
The current experimental lower bound on the mass of the SM scalar Higgs boson is about 114.5 GeV.

Now, at the one-loop level at zero temperature, the effective Higgs potential is given by
    \begin{equation}
        V(\phi, 0) = V_0 (\phi, 0) + V_1 (\phi, 0) \ ,
    \end{equation}
where the one-loop contribution $V_1(\phi, 0)$ at zero temperature is obtained
via the effective potential method as [14]
    \begin{equation}
    V_1(\phi, 0) =  \sum_l {n_l  m_l^4 (\phi) \over 64 \pi^2}
            \left [ \log \left ({m_l^2 (\phi) \over \Lambda^2 } \right ) - {3 \over 2} \right ] \ ,
    \end{equation}
where $l$ stands for various participating particles: the gauge bosons $W$, $Z$,
the third generation quarks and scalar quarks $t$, $b$, ${\tilde t}_1$, ${\tilde t}_2$,
${\tilde b}_1$ and  ${\tilde b}_2$, as well as the fourth generation quarks and scalar quarks
$t'$, $b'$, ${\tilde t'}_1$, ${\tilde t'}_2$, ${\tilde b'}_1$ and  ${\tilde b'}_2$.
The renormalization scale in the above one-loop effective potential is set as $\Lambda = m_Z$.
The degrees of freedom for each particle are:
$n_W = 6$, $n_Z = 3$, $n_t = n_b = - 12$, $n_{{\tilde t}_i} = n_{{\tilde b}_i} = 6$ ($i=1, 2$),
$n_{t'} = n_{b'} = - 12$, $n_{{\tilde t'}_i} = n_{{\tilde b'}_i} = 6$ ($i=1, 2$).
Their field-dependent masses are given by $m_W^2(\phi) = g_2^2 \phi^2/4 $,
$m_Z^2(\phi) = (g_1^2 + g_2^2) \phi^2/4$, $m_t^2(\phi) = h_t^2 \sin \beta^2 \phi^2/2$,
$m_b^2(\phi) = h_b^2 \cos \beta^2 \phi^2/2$, $m_{t'}^2(\phi) = h_{t'}^2 \sin \beta^2 \phi^2/2$,
$m_{b'}^2(\phi) = h_{b'}^2 \cos \beta^2 \phi^2/2$, and
    \begin{equation}
    m_{{\tilde q}_1 {\tilde q}_2}^2 (\phi)
    =  {m_{{\tilde q}_L}^2 (\phi) + m_{{\tilde q}_R}^2 (\phi) \over 2}
        \mp \sqrt{\left ({m_{{\tilde q}_L}^2 (\phi) - m_{{\tilde q}_R}^2 (\phi) \over 2} \right )^2
        + {\tilde A}_q^2 m_q^2 (\phi) } \ ,
    \end{equation}
with $q = t, b, t', b'$.
In the above expression for the scalar quark masses [15], we have
    \begin{eqnarray}
    m_{{\tilde t}_L}^2 (\phi) & = &
    m_Q^2 + m_t^2 (\phi) + \left ({1 \over 2}
        - {2 \over 3} \sin^2 \theta_W \right) \cos 2 \beta m_Z^2 (\phi) \ , \cr
    m_{{\tilde t}_R}^2 (\phi) & = &
    m_U^2 + m_t^2 (\phi) + {2 \over 3} \sin^2 \theta_W \cos 2 \beta m_Z^2 (\phi) \ , \cr
    m_{{\tilde b}_L}^2 (\phi) & = &
    m_Q^2 + m_b^2 (\phi) + \left (- {1 \over 2}
        + {1 \over 3} \sin^2 \theta_W \right) \cos 2 \beta m_Z^2 (\phi) \ , \cr
    m_{{\tilde b}_R}^2 (\phi) & = &
    m_D^2 + m_b^2 (\phi) - {1 \over 3} \sin^2 \theta_W \cos 2 \beta m_Z^2 (\phi) \ ,
    \end{eqnarray}
and similarly for the fourth generation by substituting with primed quantities,
where $\sin \theta_W$ is the weak mixing angle.

The parameters ${\tilde A}_t$, ${\tilde A}_b$, ${\tilde A}_{t'}$ and ${\tilde A}_{b'}$
in the above expressions for the scalar quark masses are given as
${\tilde A}_t = A_t - \mu \cot \beta$, and ${\tilde A}_b = A_b - \mu \tan \beta$,
and similarly for the fourth generation.
Note that ${{\tilde A}_t} = 0$ (${{\tilde A}_b} = 0$) does not necessarily imply that
the right-handed and the left-handed scalar top (bottom) quarks are degenerate in mass,
since there is $D$-term contributions.
Only if $D$-term contributions to the scalar top (bottom) quark masses are neglected,
${{\tilde A}_t} = 0$ (${{\tilde A}_b} = 0$) would yield degenerate right-handed and
left-handed scalar top (bottom) quarks.

However, we remark that the parameters ${\tilde A}_t$, ${\tilde A}_b$, ${\tilde A}_{t'}$
and ${\tilde A}_{b'}$ control the mixings between the scalar top or scalar bottom masses
in each generation.
If these parameters are zero, there would be no mixing between right-handed and
left-handed scalar quarks of each generation.
In this paper, we assume that there is no mixing,
taking ${\tilde A}_t = {\tilde A}_b = {\tilde A}_{t'} = {\tilde A}_{b'} =0$
in the expressions for the scalar quark masses.
Therefore, we study the MSSM Higgs potential in the decoupling limit without mixing.

The decoupling limit without mixing is an optimal situation for electroweak phase transition
to be strongly first order.
The case without mixing is more favorable for the first order electroweak phase transition
to be strong than the case with mixing, as the strength of the electroweak phase transition
is found to decrease when the mixings between scalar quarks are taken into account [15].
Also, one can notice, for example, in Fig. 2 of Ref. [16],
that the strength of the electroweak phase transition increases as $m_A$ increases
in the MSSM with three generations.
Thus, we study the MSSM Higgs potential in circumstances that enhance
the first order electroweak phase transition.

Now, the renormalized parameter $m_0^2$ in the Higgs potential can be eliminated
by the minimum condition.
By calculating the first derivative of $V (\phi, 0)$ with respect to $\phi$, $m_0^2$ is expressed as
\begin{eqnarray}
    m_0^2
    & = & { 1 \over 2} m_Z^2 \cos^2 2 \beta
    + {3 m_W^4 \over 8 \pi^2 v^2 }  \left [ \log \left ({m_W^2 \over \Lambda^2} \right ) - 1 \right ]
    + {3 m_Z^4 \over 16 \pi^2 v^2 }  \left [\log \left ({m_Z^2 \over \Lambda^2} \right ) - 1 \right ] \cr
    & &\mbox{}
    - {3 m_t^4 \over 4 \pi^2 v^2 }  \left [ \log \left ({m_t^2 \over \Lambda^2} \right ) - 1 \right ]
    - {3 m_b^4 \over 4 \pi^2 v^2 } \left [ \log \left ({m_b^2 \over \Lambda^2} \right ) - 1 \right ] \cr
    & &\mbox{}
    - {3 m_{t'}^4 \over 4 \pi^2 v^2 }  \left [ \log \left ({m_{t'}^2 \over \Lambda^2} \right ) - 1 \right ]
    - {3 m_{b'}^4 \over 4 \pi^2 v^2 } \left [ \log \left ({m_{b'}^2 \over \Lambda^2} \right ) - 1 \right ] \cr
    & &\mbox{}
    + \sum_a {3 m_a^2 \over 8 \pi^2 v^2}  (m_a^2 - m_Q^2)
    \left [ \log \left ({m_a^2 \over \Lambda^2}  \right ) - 1 \right ]  \cr
    & &\mbox{}
    + \sum_{a'} {3 m_{a'}^2 \over 8 \pi^2 v^2}  (m_{a'}^2 - m_{Q'}^2)
    \left [ \log \left ({m_{a'}^2 \over \Lambda^2}  \right ) - 1 \right ]  \ ,
\end{eqnarray}
where $a = {\tilde t}_1$, ${\tilde t}_2$, ${\tilde b}_1$, ${\tilde b}_2$, and
$a' = {\tilde t'}_1$, ${\tilde t'}_2$, ${\tilde b'}_1$, ${\tilde b'}_2$, and $v = 246$ GeV.
The mass of $h$ at the one-loop level in the decoupling limit without mixing
at zero temperature is obtained by calculating the second derivative of $V (\phi, 0)$
with respect to $\phi$ as
\begin{eqnarray}
    m_h^2
    & = & m_Z^2 \cos^2 2 \beta
    + {3 m_W^4 \over 4 \pi^2 v^2} \log \left ( {m_W^2 \over \Lambda^2}\right )
    + {3 m_Z^4 \over 8 \pi^2 v^2} \log \left ( {m_Z^2 \over \Lambda^2}\right ) \cr
    & &\mbox{}
    - {3 m_t^4 \over 2 \pi^2 v^2} \log \left ( {m_t^2 \over \Lambda^2}\right )
    - {3 m_b^4 \over 2 \pi^2 v^2} \log \left ( {m_b^2 \over \Lambda^2}\right ) \cr
    & &\mbox{}
    - {3 m_{t'}^4 \over 2 \pi^2 v^2} \log \left ( {m_{t'}^2 \over \Lambda^2}\right )
    - {3 m_{b'}^4 \over 2 \pi^2 v^2} \log \left ( {m_{b'}^2 \over \Lambda^2}\right ) \cr
    & &\mbox{}
    + \sum_a {3 (m_a^2 - m_Q^2)^2  \over 4 \pi^2 v^2} \log \left ({m_a^2 \over \Lambda^2} \right )
    + \sum_{a'} {3 (m_{a'}^2 - m_{Q'}^2)^2  \over 4 \pi^2 v^2} \log \left ({m_{a'}^2 \over \Lambda^2} \right )  \  ,
\end{eqnarray}
where $a = {\tilde t}_1$, ${\tilde t}_2$, ${\tilde b}_1$, ${\tilde b}_2$, and $a' = {\tilde t'}_1$, ${\tilde t'}_2$, ${\tilde b'}_1$, ${\tilde b'}_2$.

Now, we study the effect of finite temperature.
The one-loop contribution at finite temperature is given by [17]
\begin{equation}
    V_1 (\phi, T)
    =  \sum_l {n_l T^4 \over 2 \pi^2}
            \int_0^{\infty} dx \ x^2 \
            \log \left [1 \pm \exp{\left ( - \sqrt {x^2 + {m_l^2 (\phi)/T^2 }} \right )  } \right ] \ ,
\end{equation}
where $l = W$, $Z$, $t$, $b$, ${\tilde t}_1$, ${\tilde t}_2$, ${\tilde b}_1$ and
${\tilde b}_2$, $t'$, $b'$, ${\tilde t'}_1$, ${\tilde t'}_2$, ${\tilde b'}_1$ and  ${\tilde b'}_2$.
and the negative sign is for bosons and the positive sign for fermions.
The full one-loop effective potential at finite temperature that we are considering can
now be expressed as
\begin{equation}
        V(\phi, T) = V_0(\phi,0) + V_1(\phi, 0) + V_1(\phi, T) \ .
\end{equation}
We perform the exact integration in  $V(\phi, T)$ instead of employing the high-temperature
approximation.

\section{Numerical Analysis}

At the tree level, we have only one free parameter $\tan \beta$ in the decoupling limit
where $m_A \gg m_Z$.
At the one-loop level, the number of free parameters increases as $m_t$, $m_b$, $m_{t'}$,
$m_{b'}$, and $m_Q$, $m_{Q'}$ are introduced to the one-loop contributions.
To be concrete, we set  $m_t$ = 175 GeV and $m_b$ = 4.5 GeV, and we take for simplicity
the soft SUSY breaking parameters as $m_Q^2 = m_U^2 = m_D^2$, and similarly for the fourth generation.
Since we assume no mixing in the scalar quark sector,
we set ${\tilde A}_t = {\tilde A}_b = {\tilde A}_{t'} = {\tilde A}_{b'} = 0$.
For the masses of the fourth generation quarks, there are some experimental constraints.
Some years ago, Tevatron data have set $m_{b'} > 119$ GeV, and recently, from the search
for long-lived charged massive particles at Tevatron come more stringent experimental lower bounds
of $m_{b'} > 180$ GeV and $m_{t'} > 230$ GeV [18].
With these constraints in mind, we take $m_{t'}$ = 250 GeV and $m_{b'} = 200$ GeV.
Finally, we take 1 TeV for the value of $m_Q$ from the SUSY breaking scale $M_{\rm SUSY} = 1$ TeV.
Thus, our numerical analysis involves two free parameters: $\tan \beta$ and $m_{Q'}$.

In Fig. 1, we show a typical behavior of $V(\phi,T)$  as a function of $\phi$,
at a critical temperature $T= T_c = 84.225$ GeV, where we take $\tan \beta = 20$ and $m_{Q'} = 100$ GeV.
As one can see in the figure, we obtain the critical vacuum expectation value
as $v_c$ = 149 GeV, and the ratio as $v_c/T_c = 1.769$.
one can notice that the potential in Fig. 1 allows a strongly first-order
electroweak phase transition.
The scalar quark masses are obtained as $m_{{\tilde t}_1} = 1013$ GeV,
$m_{{\tilde t}_2} = 1014$ GeV, $m_{{\tilde b}_1} = 1000$ GeV, $m_{{\tilde b}_2} = 1001$ GeV,
$m_{{\tilde t'}_1} = 263$ GeV, $m_{{\tilde t'}_2} = 266$ GeV, $m_{{\tilde b'}_1} = 225$ GeV,
and $m_{{\tilde b'}_2} = 231$ GeV.

For the mass of $h$, we obtain $m_h$ = 129 GeV with the parameter values of Fig. 1.
This number is a little larger than the experimental lower bound on the SM Higgs boson mass, 115 GeV.
In the decoupling limit of $m_A \gg m_Z$, the behavior of $h$ is identical to that
of the SM Higgs boson.
The search for a light Higgs boson in the MSSM by DELPHI collaboration without mixing,
where only three generations of quarks are taken into account, suggests that $m_h$ is about 115 GeV,
for $\tan \beta > 15$ [19].
Comparing this number with our result of 129 GeV in Fig. 1, we may well deduce
that the difference is due to the contribution by the fourth generation of quarks.
The contribution by the fourth generation of quarks also enables the electroweak phase transition
strongly first order.
Our value is compatible with the result of Ref. [8] where the upper bound
on the lightest Higgs boson mass is obtained as 152 GeV for small
$\tan \beta$ and $96 \le m_{t'}, m_{b'} \le 125$ GeV in the MSSM with four generations of quarks.
If the mass difference between $t'$ and $b'$ is small, they may have smaller masses [9].

Now, let us study other regions of the parameter space.
We vary $m_{Q'}$ while fixing $\tan\beta =20$.
The masses of the third generation of scalar quarks are then the same as Fig. 1:
$m_{{\tilde t}_1} = 1013$ GeV, $m_{{\tilde t}_2} = 1014$ GeV, $m_{{\tilde b}_1} = 1000$ GeV,
and $m_{{\tilde b}_2} = 1001$ GeV.
The masses of the fourth generation of scalar quarks, as well as other relevant quantities,
are shown in Table I.
The second row of Table I for $m_{Q'} = 100$ GeV corresponds to the numerical result of Fig. 1.
One can see that the strength of the first order electroweak phase transition decreases
as $m_{Q'}$ increases until $m_{Q'}$ reaches 140 GeV, beyond which $v_c/T_c$ becomes less than 1.
Thus, the electroweak phase transition remains strongly first order for $m_{Q'} \le 140$ GeV.

We need to explore the boundary of the parameter space beyond which the electroweak phase transition
is no longer strongly first order.
In Table I, one can see that $v_c/T_c \sim 1$ for $\tan \beta = 20$ and $m_{Q'}$ = 140 GeV.
From this point, we examine several points of ($m_{Q'}$, $\tan \beta$) which yield
$v_c/T_c \sim 1$, by adjusting $T_c$.
For each value of $\tan\beta$, we find the upper bound value of $m_{Q'}$,
beyond which $v_c/T_c$ becomes less than 1.
The result is shown in Table II.
The fourth row of Table II is corresponds to the fourth row of Table I.
The numbers in Table II indicate that the electroweak phase transition can be
strongly first order for $ 2 \le \tan \beta \le 40$ if $m_{Q'} \le 140$ GeV.

In Fig. 2, we plot the numerical results of Table 2 on ($m_{Q'}, \tan \beta$)-plane.
The dashed curves denote the contours of $m_h$.
The solid curve denotes the contour of $v_c/T_c = 1$.
On the left-hand side of the solid curve we have $v_c/T_c \ge 1$, that is,
the electroweak phase transition is strongly first order in the region to the left of the solid curve.
Consequently, we find a region in the parameter space of the MSSM with four generations of quarks
where a strongly first order electroweak phase transition is allowed.
The parameter values of the allowed region are within the experimental constraints,
and yield $m_h$ consistent with experimental lower bound.

\begin{table}[ht]
\caption{Some values of $m_{Q'}$ for which the first order electroweak phase transition is
strong ($v_c/T_c > 1.0$).
The critical temperatures $T_c$ are obtained for which the finite temperature effective potential
has two degenerate vacua.
The relevant parameter values are the same as Fig. 1, that is, $\tan \beta = 20$, $m_Q = 1$ TeV,
$m_{t'}$ = 250 GeV, and $m_{b'} = 200$ GeV.}
\begin{center}
    \begin{tabular}{c||c|c|c|c|c|c|c|c}
    \hline
    \hline
    $m_{Q'}$ &  $m_{{\tilde t'}_1}$ & $m_{{\tilde t'}_2}$ & $m_{{\tilde b'}_1}$
    & $m_{{\tilde b'}_2}$ & $m_h$  & $T_c$ & $v_c$ & $v_c/T_c$ \\
    \hline
    \hline
    50  & 249 & 252 & 207 & 214 & 122 & 76.090 & 185 & 2.431  \\
    \hline
    100 & 263 & 266 & 225 & 231 & 129 & 84.225 & 149 & 1.769  \\
    \hline
    130 & 276 & 279 & 239 & 245 & 134 & 90.180 & 113 & 1.253  \\
    \hline
    140 & 281 & 284 & 245 & 251 & 136 & 92.268 & 93 & 1.007 \\
    \hline
    150 & 286 & 289 & 251 & 256 & 138 & 94.384 & 76 & 0.805 \\
    \hline
    \hline
    \end{tabular}
\end{center}
\end{table}

\begin{table}[ht]
\caption{Some values of ($m_{Q'}, \tan \beta$) beyond which the first order electroweak
phase transition is strong ($v_c/T_c \sim 1.0$).
The relevant parameters values are the same as Table I.
The masses of the scalar quarks of the fourth generation are calculated as about
$m_{{\tilde t'}_1} = 282$ GeV, $m_{{\tilde t'}_2} = 285$ GeV, $m_{{\tilde b'}_1} = 246$ GeV,
$m_{{\tilde b'}_2} = 251$ GeV.}
\begin{center}
    \begin{tabular}{c|c||c|c|c|c}
    \hline
    \hline
    $\tan \beta$ & $m_{Q'}$  & $m_h$ & $T_c$ & $v_c$ & $v_c/T_c$ \\
    \hline
    \hline
    2  & 147 & 120 & 87.163 & 88 & 1.009  \\
    \hline
    5  & 142 & 133 & 91.256 & 92  & 1.008 \\
    \hline
    10 & 140 & 135 & 91.965 & 92  & 1.000 \\
    \hline
    20 & 140 & 136 & 92.268 & 93  & 1.007 \\
    \hline
    30 & 140 & 137 & 92.326 & 93  & 1.007 \\
    \hline
    40 & 140 & 140 & 92.346 & 93  & 1.007 \\
    \hline
    \hline
    \end{tabular}
\end{center}
\end{table}

\section{Conclusions}

Up to now, we study the possibility of a strongly first order electroweak phase transition
in the MSSM with sequential four generations of quarks.
We assume that $m_A \gg m_Z$ and ${\tilde A}_l = 0 (l = t, b, t', b')$, that is,
we work in the decoupling limit and in the case of no mixing between scalar quarks.
We choose the relevant parameter values to be: $m_t$ = 175 GeV, $m_b$ = 4.5 GeV,
$m_{t'}$ = 250 GeV, $m_{b'} = 200$ GeV, and $m_Q = 1$ TeV. We take $m_Q^2 = m_U^2 = m_D^2$,
and similarly for the fourth generation.
These numbers are consistent with experimental constraints.

We search the parameter space of ($m_{Q'}, \tan \beta$)-plane to examine
if the electroweak phase transition is strongly first order.
We find that there are regions in the ($m_{Q'}, \tan \beta$)-plane that satisfy
our criterion of $v_c/T_c \ge 1$.
For $ 2 \le \tan \beta \le 40$,  the electroweak phase transition is strongly first order
in the region where $m_{Q'} \le 140$ GeV.
The scalar quark masses of the fourth generation are controlled mainly
by the soft SUSY breaking parameter $m_{Q'}$.
In the region where the electroweak phase transition is strongly first order,
the scalar quark masses of the fourth generation are obtained to be larger than
the quark masses of the fourth generation.

\vskip 0.3 in

\noindent
{\large {\bf Acknowledgments}}
\vskip 0.2 in
\noindent
This work was supported by Korea Research Foundation Grant (2001-050-D00005).

\vskip 0.2 in


\vfil\eject

{\bf Figure Captions}

\vskip 0.3 in
\noindent
Fig. 1: The plot of $V(\phi,T)$ as a function of $\phi$, for the critical temperature
$T_c$ = 84.225 GeV.
The relevant parameter values are set as follows: $\tan \beta = 20$, $m_Q = 1$ TeV,
$m_{t'}$ = 250 GeV, $m_{b'} = 200$ GeV, and $m_{Q'} = 100$ GeV.
One can see that the electroweak phase transition is first order.
The critical VEV is obtained as $v_c$ = 149 GeV.
Thus, $v_c/T_c = 1.769$, and the first order electroweak phase transition is strong.
The potential in this figure yields $m_h = 129$ GeV, $m_{{\tilde t}_1} = 1013$ GeV,
$m_{{\tilde t}_2} = 1014$ GeV,
$m_{{\tilde b}_1} = 1000$ GeV, $m_{{\tilde b}_2} = 1001$ GeV, $m_{{\tilde t'}_1} = 263$ GeV,
$m_{{\tilde t'}_2} = 266$ GeV, $m_{{\tilde b'}_1} = 225$ GeV, and $m_{{\tilde b'}_2} = 231$ GeV.

\vskip 0.3 in
\noindent
Fig. 2: Plots of the lightest scalar Higgs boson mass and the criterion
of the strongly first order electroweak phase transition in the ($m_{Q'}, \tan \beta$)-plane.
The remaining relevant parameters are set as $m_Q = 1$ TeV, $m_{t'}$ = 250 GeV, and $m_{b'} = 200$ GeV.
The solid curve is the contour of $v_c/T_c = 1$,
and the dashed curves are the contours of $m_h = 115$, 120, 125, 130, 135, and 140 GeV.
The masses of the scalar quarks of the third generation are calculated the same as Fig. 1,
that is, $m_{{\tilde t}_1} = 1013$ GeV, $m_{{\tilde t}_2} = 1014$ GeV, $m_{{\tilde b}_1} = 1000$ GeV,
and $m_{{\tilde b}_2} = 1001$ GeV.
The region to the left-hand side of the solid curve and above the dashed curve of $m_h = 115$ GeV
is where the first order electroweak phase transition is strong ($v_c/T_c \ge 1$) and $m_h$
is consistent with the experimental lower bound.

\setcounter{figure}{0}
\def\figurename{}{}%
\renewcommand\thefigure{Fig. 1}
\begin{figure}[t]
\epsfxsize=12cm
\hspace*{2.cm}
\epsffile{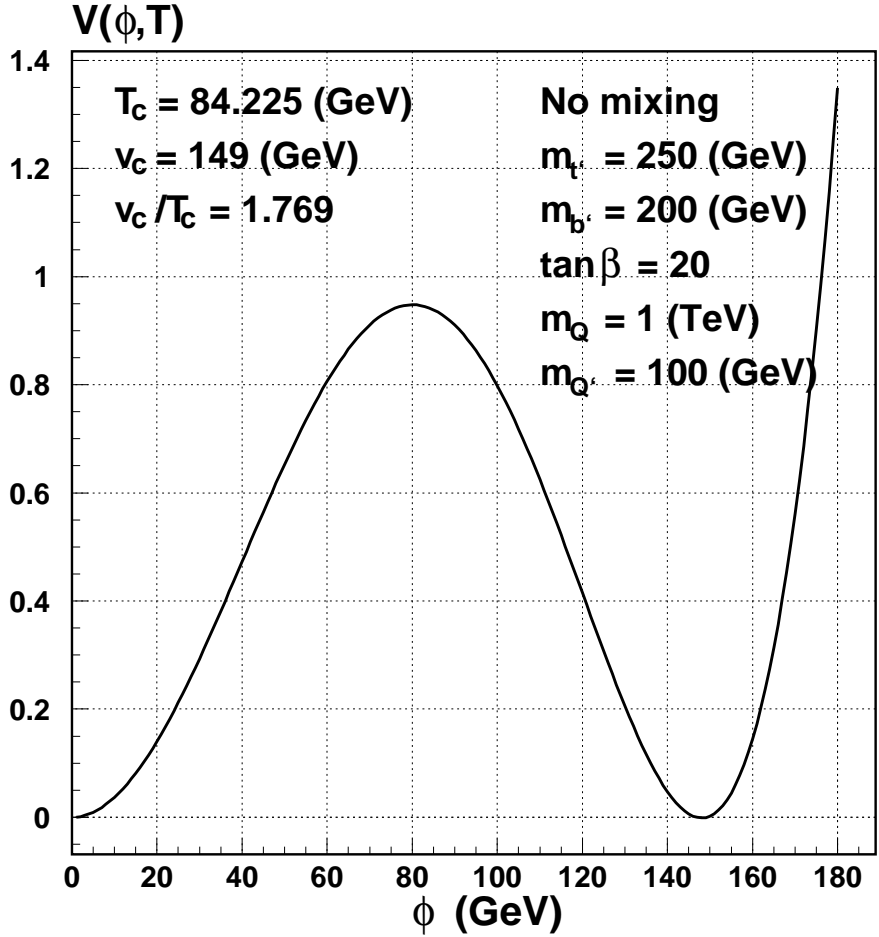}
\caption[plot]{The plot of $V(\phi,T)$ as a function of $\phi$, for the critical temperature
$T_c$ = 84.225 GeV.
The relevant parameter values are set as follows: $\tan \beta = 20$, $m_Q = 1$ TeV, $m_{t'}$ = 250 GeV,
$m_{b'} = 200$ GeV, and $m_{Q'} = 100$ GeV.
One can see that the electroweak phase transition is first order.
The critical VEV is obtained as $v_c$ = 149 GeV.
Thus, $v_c/T_c = 1.769$, and the first order electroweak phase transition is strong.
The potential in this figure yields $m_h = 129$ GeV, $m_{{\tilde t}_1} = 1013$ GeV,
$m_{{\tilde t}_2} = 1014$ GeV,
$m_{{\tilde b}_1} = 1000$ GeV, $m_{{\tilde b}_2} = 1001$ GeV, $m_{{\tilde t'}_1} = 263$ GeV,
$m_{{\tilde t'}_2} = 266$ GeV, $m_{{\tilde b'}_1} = 225$ GeV, and $m_{{\tilde b'}_2} = 231$ GeV.}
\end{figure}
\renewcommand\thefigure{Fig. 2}
\begin{figure}[t]
\epsfxsize=12cm
\hspace*{2.cm}
\epsffile{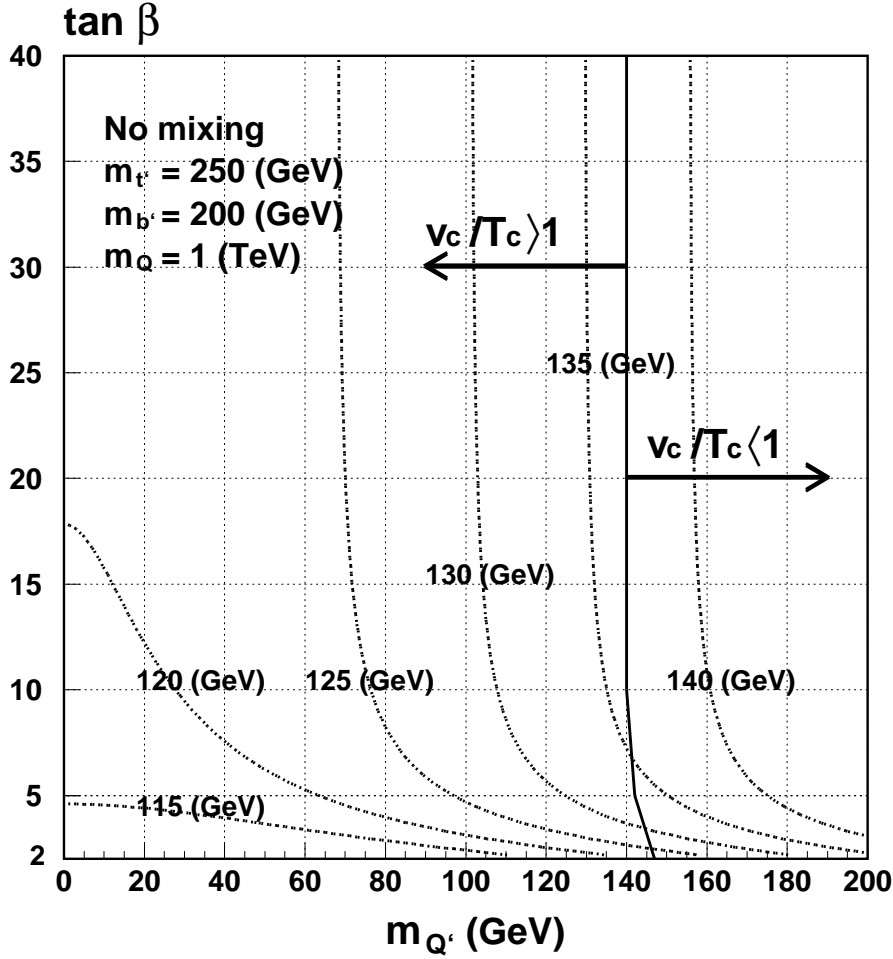}
\caption[plot]{Plots of the lightest scalar Higgs boson mass and the criterion
of the strongly first order electroweak phase transition in the ($m_{Q'}, \tan \beta$)-plane.
The remaining relevant parameters are set as $m_Q = 1$ TeV, $m_{t'}$ = 250 GeV,
and $m_{b'} = 200$ GeV.
The solid curve is the contour of $v_c/T_c = 1$,  and the dashed curves are the contours of
$m_h = 115$, 120, 125, 130, 135, and 140 GeV.
The masses of the scalar quarks of the third generation are calculated the same as Fig. 1,
that is, $m_{{\tilde t}_1} = 1013$ GeV, $m_{{\tilde t}_2} = 1014$ GeV, $m_{{\tilde b}_1} = 1000$ GeV,
and $m_{{\tilde b}_2} = 1001$ GeV.
The region to the left-hand side of the solid curve and above the dashed curve of $m_h = 115$ GeV
is where the first order electroweak phase transition is strong ($v_c/T_c \ge 1$) and $m_h$
is consistent with the experimental lower bound.}
\end{figure}
\end{document}